\documentclass[showpacs,twocolumn,preprintnumbers,footinbib,pre,amsmath,amssymb]{revtex4}
\usepackage{graphicx}% Include figure files
\usepackage{dcolumn}% Align table columns on decimal point
\usepackage{bm}% bold math

\begin{document}

\preprint{Phys.Rev.E,79(1), in press.}

\title{A noise-driven attractor switching device}

\author{Naoki Asakawa}

\thanks{Corresponding authors: email: asakawa@sanken.osaka-u.ac.jp, kawai@sanken.osaka-u.ac.jp}
\affiliation{Institute of Scientific and Industrial Research, Osaka University,
Mihogaoka, Ibaraki, Osaka, Japan}

\author{Yasushi Hotta}
\affiliation{Institute of Scientific and Industrial Research, Osaka University,
Mihogaoka, Ibaraki, Osaka, Japan}

\author{Teruo Kanki}
\affiliation{Institute of Scientific and Industrial Research, Osaka University,
Mihogaoka, Ibaraki, Osaka, Japan}

\author{Hitoshi Tabata}
\affiliation{Graduate School of Engineering, University of Tokyo, Hongo, Bunkyo-ku, Tokyo, Japan}

\author{Tomoji Kawai}
\thanks{Corresponding authors: email: asakawa@sanken.osaka-u.ac.jp, kawai@sanken.osaka-u.ac.jp}
\affiliation{Institute of Scientific and Industrial Research, Osaka University,
Mihogaoka, Ibaraki, Osaka, Japan}

\date{\today}% It is always \today, today,


\begin{abstract}
Problems with artificial neural networks originate from                        %58  
their deterministic nature and inevitable prior learnings, resulting in          %72 
inadequate adaptability against unpredictable, abrupt environmental change.       %75 
Here we show that a stochastically excitable threshold unit can be utilized by these systems          %91
to partially overcome the environmental change.                                              %36 %47
Using an excitable threshold system,						 %34 
attractors were created								 %24 
that represent quasi-equilibrium states into which a system settles		 %68 
until disrupted by environmental change.					 %40 
Furthermore,									 %13 
noise-driven attractor stabilization and switching were embodied		 %65 
by inhibitory connections.							 %27 
Noise works as a power source to stabilize and switch attractors, and		 %70 
endows the system with hysteresis behavior that resembles that of stereopsis and %81 
binocular rivalry in the human visual cortex.                                        %41 
A canonical model of the ring network with inhibitory connections composed of class 1 neurons %94
also shows properties that are similar to the simple threshold system.                  %63
\end{abstract}

\maketitle

\newpage
\section{introduction}
Sensory nerves of several biological systems including 
crickets\cite{cricket}, crayfishes\cite{crayfish} and paddlefishes\cite{paddlefish}
exploit the phenomenon of stochastic resonance (SR) 
for weak signal detection within a noisy environment\cite{SR}. 
Previous studies concerning cerebral physiology have provided evidence 
of SR following analysis of hippocampal slices from rat\cite{rat_hippocampus} and human\cite{human_brain1,human_brain2} brains.
For an adaptive system to successfully operate in an environment,
it would be preferable to utilize a certain number of attractors,
which represent quasi-equilibrium states into which the system settles until disrupted by environmental change, 
and an autonomous selection mechanism for one appropriate attractor in the environment.
However, the conventional view of SR phenomena lacks the concept that a process switches stochastically  
to the most preferable attractor.
Recently, 
the concept of neuronal computations with stochastic network states in the brain
has also been accepted\cite{neuronal_computation}.
However, the functional role of noise in neuronal activity remains unknown.
In this article,
we assembled attractors composed of reverberating electronic circuits
in order to investigate the functional role of noise.
The results obtained using an SR system based on an excitable threshold system\cite{MP} revealed a higher-order function of the SR system and noise-driven attractor switching with a hysteresis phenomenon, 
which is beyond the well-known functions of noise-driven enhancement 
on both sensitivity to weak signal detection and signal propagation.
In order to examine whether the higher-order function can be extended to more general neuronal models
than the simple excitable threshold system, 
we also performed a numerical simulation for a canonical model of a class 1 neural network\cite{Ermentrout,Ermentrout2,Izhikevich1999,Izhikevich2,Kanamaru1,Kanamaru2,Kanamaru3}.

\section{Methods}
\subsection{A simple excitable threshold unit by an analog electronic circuit}
\begin{figure}[htbp]
\includegraphics[scale=0.5]{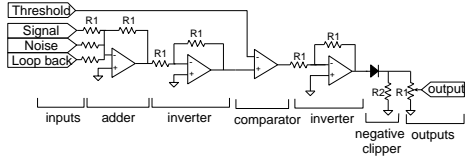}
\caption{Analog electric circuit diagram of a simple excitable threshold unit (McCulloch-Pitts neuron\cite{MP}).
The unit consists of external signal inputs including sensory signals and
signals from the other modules or rings,
external noise input,
a summation of these components, 
a successive comparator and an output port.
The negative clipper is inserted between the comparator and output port
in order to insure that an output with asymmetric voltage is obtained.
$R_1$ and $R_2$ are 10 k$\Omega$ and 100 k$\Omega$, respectively.
We used quad operational amplifiers, TL084 (Texas Instruments, inc.).
The positive and negative electric power source used for the amplifiers was $\pm$ 12 V.
}
\label{Fig1}
\end{figure}

Fig. \ref{Fig1} shows an example of an analog operational amplifier circuit of 
the simple excitable threshold unit.
The circuit is composed of seven functional parts:
multiple signal inputs, 
adder, 
inverter, 
comparator,
inverter, 
negative clipper, 
and an output port.
The negative feedback circuits of operational amplifiers with a gain of unity 
were used for the adder and inverters.
We used quad operational amplifiers, TL084 (Texas Instruments, Inc.).
The positive and negative electric power source used for the amplifiers was $\pm$ 12 V.
The diode for the negative clipper was 1S1558 (Toshiba, Corp., Japan).
All the resistors used here are standard metal-film resistors.

\subsection{Implementation of an inhibitory connection by an analog electronic circuit}
Localization of neuronal excitations is a basic feature of brain function.
In order to select an appropriate attractor in a certain environment,
inhibitory connections are crucial to the localization of neuronal excitations.
It would also be important for the connection to realize attractor switching against environmental change.
An inhibitory connection was realized using the variable threshold of the excitable unit. 
The shutdown of noise generators that is triggered by the output of the other units can also be 
used as an inhibitory connection.
Alternatively, the signal input of the inverted output of the other unit 
will be available for this purpose.

\subsection{Electronic circuit experiments}
All data of output voltage for the attractor switching electronic circuit (Figs. \ref{Fig1} and \ref{Fig3})
were collected by a multifunction data logger (USB-6008; 12-bit longitudinal digital resolution; maximum sampling rate of 10 kS/sec, National Instruments, Corp.),
which was controlled through a general purpose interface bus (GPIB) via a high-speed universal serial bus (USB 2.0). 
The GPIB and data logger were handled using a graphical user interface built using LabVIEW 8.2 (National Instruments, Corp.).
Input signals and threshold voltages were synthesized by standard function generators (WF1974, NF, Corp., Japan).
Gaussian white noise was synthesized using a function generator (33120A, Agilent Technologies, Inc.) and 
fed into the input port of the simple excitable units 
through an eighth-order Butterworth low-pass filter with a cutoff frequency of 50kHz(two VT-4BLA in box \#3334, NF, Corp., Japan).
The sampling rate of data collection was 1 k/sec.
The average firing rate was determined by averaging digitized output signals during 0.1 s (100 points).
The response to the sinusoidal input signal (0.1 V$\times \sin(\omega t)$; $\omega/2\pi$ = 0.3 Hz) was recorded for 18 cycles (only data for the first three cycles is shown in Fig. \ref{Fig3}b).
The parameters used in Fig. \ref{Fig3}b were shown in Table.\ref{table1}.

\subsection{Numerical simulation for the network of simple excitable units}
The numerical simulation for simple excitable threshold systems was performed using McCulloch-Pitts neurons\cite{MP}.
Let us consider a ring circuit (Fig. \ref{Fig2}c) of $M$ modules(Fig. \ref{Fig2}b), 
where each module is composed of $N$ parallel neurons(Fig. \ref{Fig2}a).
The output of the $m$-th module with $N$ parallel neurons in ring $X$, $V^X_m(t)$, at time $t$ 
is calculated using the following equation:
\begin{eqnarray}
V^X_m(t) &=& \frac{1}{N} \sum^{N}_{n=1} {\cal H}[x^X_{mn}(t) - V^{\rm Th}_X(t)]\\
%& & \left\{ \begin{array}{ll}
%x^X_{\rm 1n}(t) = x^{\rm s}_{1n}(t) + \epsilon^X_{\rm M} V^X_{\rm M}(t-\tau) + \xi^X_{1n}(t) & \mbox{if $m=1$}\\
%x^X_{\rm mn}(t) = \epsilon^X_{\rm m-1} V^X_{\rm m-1}(t-\tau) + \xi^X_{mn}(t) & \mbox{if $1<m \le M$},
& & \left\{ \begin{array}{l}
x^X_{\rm 1n}(t) = x^{\rm s}_{1n}(t) + \epsilon^X_{\rm M} V^X_{\rm M}(t-\tau) + \xi^X_{1n}(t)\\
\quad \quad \quad \quad \quad \quad \quad \quad \quad \mbox{if $m=1$}\\
x^X_{\rm mn}(t) = \epsilon^X_{\rm m-1} V^X_{\rm m-1}(t-\tau) + \xi^X_{mn}(t)\\
\quad \quad \quad \quad \quad \quad  \quad \quad \quad \mbox{if $1<m \le M$},
\end{array}
\right.
\end{eqnarray}
where ${\cal H}[u]$ is a Heaviside step function:
\begin{eqnarray}
{\cal H}[u] &=& \left\{ \begin{array}{ll}
			1 & \mbox{if $u > 0$}\\
			0 & \mbox{otherwise}.
			\end{array}
		\right.
\end{eqnarray}
$V^{\rm Th}_X(t)$, $x^{\rm s}_{mn}(t)$, $\xi^X_{mn}(t)$ and $\tau$ are
the threshold value, sensory signal input, noise for the $n$-th excitable unit in the $m$-th module, and delay time, respectively.
For simplicity, all the threshold units in a ring $X$ have the same threshold value($V^{\rm Th}_X(t)$).
Note that the delay time is important to realize the recurrent network.
$\epsilon^X_m$ is
the coupling constant between the $m$-th and $(m+1)$-th modules and 
is defined as the ratio of the input amplitude of $(m+1)$-module to the output amplitude of the $m$-th module.
Noise is represented by the mutually independent and uncorrelated functions 
\begin{eqnarray}
\langle \xi^X_{mn}(t) \xi^X_{pq}(t') \rangle &=&  D \delta_{mp} \delta_{nq} \delta(t-t')\\
\langle \xi^X_{mn}(t) \rangle &=& 0.
\end{eqnarray}
where $\langle \cdot \rangle$ denotes time autocorrelation.
The distribution function, $p(x)$, with $\xi^X_{mn}(t)=x$ is defined as follows:
\begin{eqnarray}
p(x) &=& \left\{ \begin{array}{ll}
\mbox{const.} & \mbox{:uniform noise}\\
\frac{1}{\sqrt{\pi D}} \exp(-\frac{x^2}{D}) & \mbox{:Gaussian noise}\\
\end{array}
\right.
\end{eqnarray}

The numerical simulations for simple excitable threshold systems were performed on a personal computer with a Core 2 processor (1.86 GHz) and 1 GB of random access memory.
A simulation program was written using GNU Octave (version.2.1.73) under the Vine Linux OS (version 3.3.6).
All the parameters used in the simulation were dimensionless and summarized in Table. \ref{table1}.

\subsection{Numerical simulation for the ring network with asymmetric inhibitory connections using a canonical model for class 1 neurons}
A canonical model of class 1 neurons was employed to further investigate the properties of 
a two-ring system of stochastically excitable threshold units with inhibitory connections (vide infra). 
Based on the theory by Ermentrout\cite{Ermentrout,Ermentrout2}, 
Izhikevich\cite{Izhikevich1999,Izhikevich2}, and Kanamaru\cite{Kanamaru1,Kanamaru3},
we interpreted the Langevin equation in the Stratonovich's sense:
\begin{eqnarray}
& &\dot{\theta}^X_{mn}(t) = (1-\cos \theta^X_{mn}(t)) + (1+\cos \theta^X_{mn}(t)) \nonumber\\
& & \ \ \ \ \times(r_X + \xi^X_{mn}(t) + K^{{\rm ring},X}_m(t) \nonumber\\
& & \ \ \ \ \ \ \ \ \ + K^{{\rm inh},X}_m(t) + K^{{\rm stim},X}_m(t) ), \label{phase_model}
\end{eqnarray}
where $X=A$ or $B$ and $\theta_{mn}^X(t)$, $r_X$ and $\xi_{mn}^X(t)$ are
internal states of the $n$-th neuron in the $m$-th module, the bifurcation parameter for ring $X$, and noise applied for
the $n$-th neuron in the $m$-th module, respectively.
$\xi_{mn}^X(t)$ is uniform white noise with a correlation of 
\begin{eqnarray}
\langle \xi_{mn}^X(t) \xi_{m'n'}^Y(t') \rangle = D\delta_{XY}\delta_{mm'}\delta_{nn'}\delta(t-t'),
\end{eqnarray}
where $m,m'=1,\cdots,M$ and $n,n'=1,\cdots,N$.
$\delta_{XY}$, $\delta_{mm'}$, and $\delta_{nn'}$ are Kronecker delta and 
$\delta(t-t')$ is Dirac's delta function.
$D$ is the noise amplitude.
$K_m^{{\rm ring},X}(t)$, $K_m^{{\rm inh},X}(t)$ and $K_m^{{\rm stim},X}(t)$ are
signal inputs from the module of the upper stream in the ring,
inhibitory signals from ring B to ring A,
and the external sinusoidal input, respectively.
\begin{eqnarray}
\left\{\begin{array}{l} 
K_m^{{\rm ring},X}(t) = \epsilon_{\rm ex}^X I_{m-1}^X(t) {\rm \ \ \ \ (if\  m\  \neq \  1)} \\
K_{m=1}^{{\rm ring},X}(t) = \epsilon_{\rm ex}^X I_{M}^X(t),\\
\end{array}
\right.
\end{eqnarray}
where $\epsilon_{ex}^X$ is the excitatory coupling constant between modules in ring $X$.
The asymmetric inhibitory connection is expressed as follows:
\begin{eqnarray}
\left\{\begin{array}{l}
K_m^{{\rm inh},A}(t) = - \ \epsilon_{\rm inh} \sum^{M}_{m'=1}I_{m'}^B(t)\\
K_m^{{\rm inh},B}(t) = 0,\\
\end{array}
\right.
\end{eqnarray}
where output current of the $m$-th module, $I_m^X(t)$, is expressed as an averaging over individual output currents, $I_{mn}^X(t)$, of the $n$-th neuron in the $m$-th module of ring X because of the parallel structure of modules
(see Fig. \ref{Fig2}b),
\begin{eqnarray}
I_m^X(t)&=&\frac{1}{N}\sum^N_{n=1}I_{mn}^X(t) \label{sum_Im}\\
\dot{I_{mn}^X(t)}&=&-\frac{1}{\kappa_X}(I_{mn}^X(t) - \delta(\theta_{mn}^X - \pi)), \label{decay_Imn}
\end{eqnarray}
and $\epsilon_{\rm inh}$ is the inhibitory coupling constant between rings.
A sinusoidal sensory input is applied to the first module ($m=1$) in each ring. 
\begin{eqnarray}
K_m^{{\rm stim},X}(t) = A_{\rm sig}\sin(\omega t) {\rm \ \ \ \ (if\  m=1)},
\end{eqnarray}
where $A_{\rm sig}$ and $\omega$ are the amplitude and angular frequency of the sinusoidal input signal, respectively. 
From Eq.\ref{phase_model}, \ref{sum_Im} and \ref{decay_Imn}, 
the dynamics of $\theta_{mn}^X$ was numerically calculated 
using an ordinary differential equation solver, "LSODE"\cite{LSODE}.
All the parameters in the simulation were given in Table. \ref{table2}.

\section{Results and Discussion}
We focus on 
a simplified version of a recurrent neural network: a ring circuit.
In this architecture, positive feedback is localized to a single ring, 
which is the minimum component of a complex web of the neural network.
We first assembled a memory ring circuit driven by noise.
The ring circuit shows a fundamental property of positive feedback 
that is found in the hierarchy of complex information processing in the human brain\cite{Amit, Iyengar}.
We explored the properties of a single-ring circuit (Fig. \ref{Fig2}c) of 
$M$ modules (Fig. \ref{Fig2}b), where each module was composed of $N$ parallel threshold excitable units (Fig. \ref{Fig2}a).
This module is the so-called Collins model\cite{Collins} with time delay,
and it enhances the system's sensitivity to weak inputs 
without adjusting the non-zero level of noise amplitude,
leading to enhancement of weak signal detection and 
therefore signal propagation in sensory neurons.
The numerical simulation for simple excitable threshold systems was performed using 
McCulloch-Pitts neurons\cite{MP}
with uniform white noise, and dimensionless parameters were used to maintain the generality of the problem. 
The threshold and output amplitudes are defined as $V^{\rm Th}_m$ and $V_m$, respectively, and
the details of the simulation are shown in Sec.II.D and the caption of Fig. \ref{Fig2}.
We looked at the stochastic response of the ring as a function of 
the number of parallels ($N$) in each module(Fig. \ref{Fig2}d).
In Fig. \ref{Fig2}d, 
the pulsed sinusoidal signal is stored in the ring
as a form of summed neuronal spikes 
if independent noise with appropriate amplitude is applied to each unit.
An extended lifetime of the stored signal was observed 
with increasing $N$ (Fig. \ref{Fig2}d).

In addition to the structural factor of ring circuits,
we found that the noise amplitude and inter-module coupling constant 
were important in determining the decay rate of transient memory storage.
Phase diagrams of the decay rate of memory (Fig. \ref{Fig2}e) and 
signal-to-noise ratio (SNR) of the peak at a frequency of sinusoidal input in the Fourier power spectrum (Fig. \ref{Fig2}f) 
were obtained by numerical simulation (see Sec.II.D) as functions of the inter-module coupling constant ($\epsilon$) 
and noise amplitude ($D$) 
for the single ring of $(M,N)$ = (4,100).
The rate was determined from the full-width at half-maximum ($\Delta \omega$) 
of the peak at the dimensionless input frequency ($\omega/2\pi=6.25\times10^{-2}$) 
of a fast Fourier power spectrum, and
the value of SNR was defined as the ratio of the peak intensity with the averaged noise intensity 
at 100 points of the proximate baseline over the higher frequency than $\omega + \Delta \omega$. 
A transient memory domain appears between the non-firing domain ("NF") and the reverberating domain ("R") (Fig. \ref{Fig2}e).
The SNR under conditions of a noise amplitude comparable to the perithreshold signal tends to
be largely irrelevant with regard to the coupling constant when $\epsilon > 0.05$ (Fig. \ref{Fig2}f),
and $\epsilon$-slice (slice at the constant $\epsilon(0.05< \epsilon <0.20)$) 
of Fig. \ref{Fig2}f indicates the typical property of stochastic resonance,
where application of noise with non-zero amplitude enhances SNR of the spectrum\cite{SR}.
Apparently, variation of the decay rate depended on the noise amplitude,
suggesting that noise plays an important role for signal transmission in the ring and
that the various energetic states of the reverberating circuit can be defined using the firing rate,
leading to the creation and stabilization of attractors (vide infra).

Taking into account the behavior of transient memory storage of the ring, we designed 
an attractor switching electronic circuit 
integrated with inhibitory connections (Fig. \ref{Fig3}a).
Inhibitory connections are important in various neural functions such as 
binocular rivalry in the visual cortex\cite{Wilson} and dynamical encoding of an olfactory system\cite{Laurent}.
In particular, 
the spatio-temporal dynamics of the locust olfactory system are important for the coding of odor intensity and identity\cite{Stopfer}. 
Our device is composed of two ring circuits with excitable modules.
The output of ring B ($\epsilon V^{\rm B}_m=0.25$V) is inhibitory and connected to ring A 
via feeding into the threshold setting port of ring A
(Fig. \ref{Fig3}a; see also Sections II.A and II.B for details), 
insuring that either ring A or B can fire.
From the electronic circuit experiment and numerical simulation, 
the behavior of transient memory storage brings about a switching effect in the device 
(Figs. \ref{Fig3}b and \ref{Fig4}a--c).
The default threshold value of ring A ($V_{\rm A}^{\rm Th}=0.03$V for the experiment and 0.05 for the simulation) was set much lower than that of ring B ($V_{\rm B}^{\rm Th}=0.06$V for the experiment and 0.20 for the simulation),
which forces ring A to be more likely to fire than ring B.
If the input signal amplitude or mean pulse input rate is large enough to fire ring B,
then output A is depressed.
Even if the input signal decreases to subthreshold for ring B,
ring B reverberates for a while due to its transient nature.
But ring A is disinhibited after the rest of ring B.
In such a system, 
we can define a two-dimensional attractor as ($R_{\rm A}$, $R_{\rm B}$) in the phase space,
where $R_{\rm A}$ and $R_{\rm B}$ are the firing rates of rings A and B, respectively.

The application of noise with an appropriate amplitude endows 
the electronic circuit with a hysteresis effect (Fig. \ref{Fig3}b), and
the numerical simulation of the device qualitatively reproduced this experimental tendency (Fig. \ref{Fig4}a). 
Interestingly,
noise with an appropriate amplitude (see the curves for $D=2.2$ and $2.4$V in Fig. \ref{Fig3}b and for $D=0.11$ and $0.12$ in Fig. \ref{Fig4}a) is required for the device to show a clear hysteresis curve.
The hysteresis effect is due to
the noise-driven transient memory storage effect of ring B and the asymmetric inhibitory connection.
During the down-sweeping process of the input signal amplitude,
the switching voltage becomes much lower than $V_{\rm B}^{\rm Th}$ because of
the above-mentioned reverberation of ring B.
During the up-sweeping process, the onset voltage of the firing of ring B 
can be determined only by $V_{\rm B}^{\rm Th}$ and the noise amplitude,
giving a higher switching voltage than that of the down-sweeping process.
Thus, the ring network circuit of the excitable threshold system with inhibitory connections
allows attractor switching with a hysteresis phenomenon.
Additionally,
the number of $N$ greatly affected the hysteresis (Fig. \ref{Fig4}b).
A device with the larger $N$ produced the larger coercivity in the hysteresis loop,
indicating that the device would 
be tied to past memory rather than new environmental sensory input.
Furthermore, the device with larger $N$ behaved more deterministically
and less probabilistically, and vice versa.
From the above discussion, 
we found the higher-order function that originated from SR; that is,
noise-driven signal propagation causes the transient memory effect of the ring, which in turn provides the function of
a noise-driven attractor switching effect with hysteresis.

In order to investigate whether the above-mentioned property of the two-ring system with asymmetric inhibition can be
generalized to other neuronal models,
we performed the similar simulation using a canonical model of a class 1 neural network\cite{Ermentrout,Izhikevich1999,Kanamaru2}. 
Details of the conditions for the simulation are given in the caption of Fig. \ref{Fig6} and Sec.II.E.
Figs. \ref{Fig6}a-c show the variation of dependence of mean firing rate for the fourth module ($m=4$) of rings A and B 
on noise amplitude ($D$), the number of parallels ($N$) in a module, and input frequency ($\omega$), respectively
(see Fig. \ref{Fig5} for device architecture). 
The qualitative tendencies with respect to $D$, $N$ and $\omega$ 
were the same as those for the simple threshold system described above (Fig. \ref{Fig4}); that is, i) noise with proper amplitude is required for the device to show clear hysteresis, and 
ii) the larger the number of $N$ and input frequency, the larger the coercivity of hysteresis,
and iii) the moderate sweep frequency is required to draw clear hysteresis.

The results shown by the simple threshold system and the canonical model of a class 1 neural network are phenomenologically consistent with recent results showing hysteresis effects in stereopsis 
and binocular rivalry in humans\cite{Wilson_hysteresis}.
In the literature,
hysteresis is measured as a function of orientation disparity in tilted gratings in which a transition is perceived
between stereopsis and binocular rivalry.
A fast tilting rate, which corresponds to high-frequency input in our case, 
results in enhancement of coercivity.
Furthermore, 
a smaller optical Michelson contrast of a sinusoidal grating causes a larger coercivity.
A previous functional magnetic resonance imaging study concerning a general mechanism for
perceptual decision-making indicated that
several brain regions associated with the attentional network show a greater response
when the task becomes more difficult\cite{decision_making}.
Since a test with a smaller grating contrast involves a more difficult task,
the region requires more attentional resources for correct performance;
that is, the number of neurons (corresponding to $N$) associated with the task could be larger.
This is consistent with our result that the device with larger $N$ gives larger coercivity.

The current type of stochastically excitable threshold device in our study essentially requires 
two parts:
a noise generator and comparator.
If these components are represented by novel materials beyond silicon materials,
novel bio-inspired devices will be realized. 
Critical dynamics near the phase transition point at ambient temperature 
will be available to the noise generator if the dynamics are coupled to physical properties 
such as electric conductivity.
Furthermore,
field-induced phase transition will be of practical importance in realizing the comparator.
Future development on suitable materials for the comparator with a threshold voltage 
much lower than that of conventional analog operational amplifiers (namely, conventional transistors) will yield 
attractor switching devices with ultra-low energy consumption like the brain.

\section{Conclusion}
Problems with artificial neural networks originate from                        
their deterministic nature and inevitable prior learnings, resulting in           
inadequate adaptability against unpredictable, abrupt environmental change.        
In this article 
we showed that a stochastically excitable threshold unit can be utilized by these systems         
to partially overcome the environmental change; that is,
timing of attractor switching is varied depending on not only
on the sensory input but also on the memory stored in the ring.
Noise-driven attractor stabilization and switching were embodied		 
by a multiple ring network with inhibitory connections.							 
Noise works as a power source to stabilize and switch attractors, and		 
endows the system with hysteresis behavior that resembles that of stereopsis and  
binocular rivalry in the human visual cortex.                                     
In order to further investigate the property of the multiple ring network with inhibitory connections,
we performed the numerical simulation of a canonical model of class 1 neurons, 
where the result showed properties that were similar to the simple threshold system.                  
The device shown in this article will also inspire the material science and engineering 
in fabrication of noise generators and comparators using functional materials with
critical dynamics near the phase transition at ambient temperature and/or 
field-induced phase transition.

\section{Acknowledgments}
This research was supported by ``Special Coordination Funds for 
Promoting Science and Technology: Yuragi Project" of the Ministry of 
Education, Culture, Sports, Science and Technology, Japan.

\section{Appendix}
\subsection{Parameters for numerical simulation and electric circuit experiments}
\begin{figure}[htbp]
\includegraphics[scale=0.5]{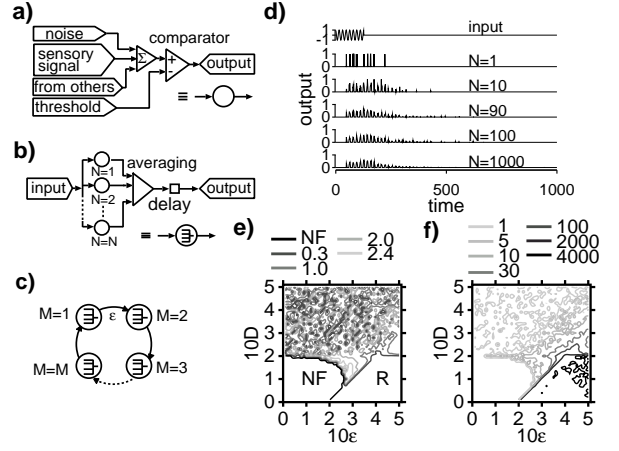}
\caption{
(a) An excitable unit is composed of the following four parts:
signal input ports
(an external sensory input, 
the signals from the other units,
 and external noise),
summation of these input signals,
a successive comparator,
and an output port.
The dimensionless output amplitude of the comparator is set to unity.
(b) A summing parallel network module (number of units: $N$) 
of the excitable units of (a), with time delay.
This is a version of the so-called Collins-type excitable network\cite{Collins} with time delay.
The input signal is fed into the input port of each unit.
Signal averaging is performed prior to the delay.
(c) A ring circuit consists of $M$ directionally coupled modules (b). 
The value $\epsilon$ denotes the coupling constant between rings, 
which scales the output voltage of the rings.
(d) Variation of transients of the output signal of the ring circuit 
of $M = 4$ as a function of the number of parallels in a module, $N$.
The other parameters used for the simulation were as follows and summarized in Table \ref{table1} in Appendix:
$\epsilon V_m$ ($V_m$: output amplitude): 0.16, 
threshold amplitude of comparators ($V^{\rm Th}_m$): 0.10,
noise amplitude ($D$): 0.10, 
signal delay at each excitable unit: 16,
input signal function: $\cos(2\pi t/16)$,
input pulse width($t_{\rm pw}$): 128,
and time-step for all simulations: 1.
(e,f) Simulated ($\epsilon,D$) phase diagrams for relaxation rate (e) and signal-to-noise ratio (SNR) (f) for the ring with ($M,N$) = (4,100) are shown.
$V^{\rm Th}_m$ was 0.20.
``NF" at the scale bar of (e) stands for ``non-firing." Note that the scale bar of (f) is nonlinear.
}
\label{Fig2}
\end{figure}

\begin{figure}[htbp]
\includegraphics[scale=0.5]{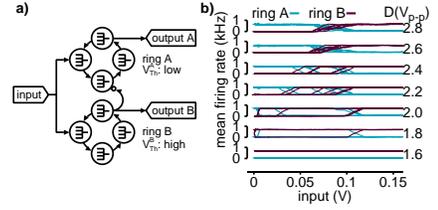}
\caption{
(Color online) (a) A circuit diagram of the attractor switching device. 
A detailed electric circuit design is presented in Sec.II.A.
(b) The experimental result of the device with ($M,N$) = (4,1) rings.
The firing rates of the fourth module in both rings was monitored as functions of 
the input signal and noise amplitudes.
The parameters for the experiment were as follows:
$\epsilon V^{\rm A}_m$: 0.16 V,
$\epsilon V^{\rm B}_m$: 0.25 V,
$V^{\rm Th}_{\rm A}$: 0.03 V,
and $V^{\rm Th}_{\rm_B}$: 0.06 V.
$D$ was varied from 1.6 V to 2.8 V with a step of 0.2 V.
See Sec.II.C for the details of the experiments.
}
\label{Fig3}
\end{figure}

\begin{figure}[htbp]
\includegraphics[scale=0.5]{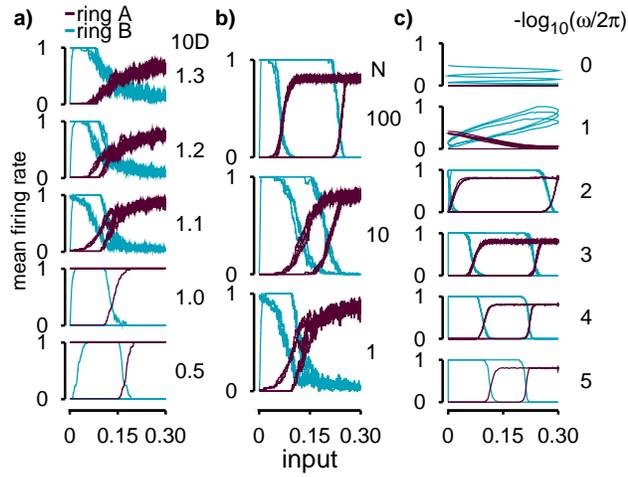}
\caption{
(Color online) (a) The simulated mean firing rate for the devices with ($M,N$) = (4,1).
The parameters for the simulation were as follows:
$V^{\rm Th}_{\rm_A}$: 0.05,
$V^{\rm Th}_{\rm B}$: 0.20,
$\epsilon V^{\rm A}_m$: 0.30,
and $\epsilon V^{\rm B}_m$: 0.30.
$D$ was set to 0.05, 0.10, 0.11, 0.12, and 0.13.
The other parameters were set to values as indicated in Fig. \ref{Fig2} and also in Table \ref{table1} in Appendix.
The inhibitory connection was implemented by varying the threshold amplitude:
$V^{\rm Th}_{\rm_A}$ = 0.05 at the disinhibition state and 2.00 at the inhibition state.
(b) The response of the device was explored as a function of $N$ while $D$ was fixed at 0.11.
(c) The frequency response of the device with $N=100$ and $D=0.11$ is shown. 
The moderate sweep frequency was necessary for hysteresis behaviors.
}
\label{Fig4}
\end{figure}

\begin{figure}[htbp]
\begin{center}
\includegraphics[scale=0.5]{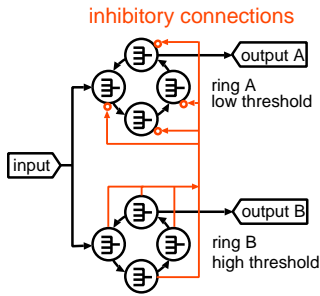}
\end{center}
\caption{
(Color online) A circuit diagram of the attractor switching device.
The output signals of all the modules in ring B were fed into the signal input ports of all the modules of ring A.
}
\label{Fig5}
\end{figure}

Here we provide all the parameters for the numerical simulations.
For the simulation of simple threshold units,
the parameters for the simulations in Figs. \ref{Fig2}d-f and \ref{Fig4}a-c
were tabulated in Table\ref{table1}.

\begin{figure}[htbp]
\includegraphics[scale=0.5]{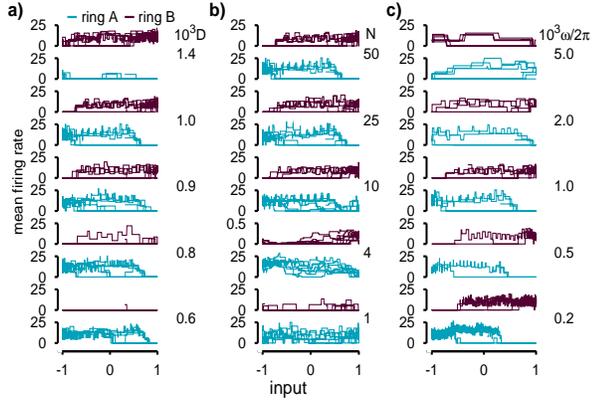}
\caption{
(Color online) The simulated mean firing rate for the canonical model of class 1 neurons as a function of noise amplitude ($D$)(a), 
the number of $N$(b), and input frequency ($\omega$)(c).
$(M,N)$=(4,4), $r_A$=-0.008, and $r_B$=-0.020.
The excitatory coupling constants ($\epsilon_{\rm ex}^A$ and $\epsilon_{\rm ex}^B$) were 0.5 and
the inhibitory connection ($\epsilon_{\rm inh}$) from ring B to ring A was 3.0. 
For a) and b), the input sinusoidal signal is 0.019$\times$ $\sin(\omega t)$, where $\omega /2 \pi$=1.0$\times$10$^{-3}$.
For b) and c), $D$=1.0$\times$10$^{-4}$.
For a) and c), $N$=50
The network is composed of slowly connected class 1 neurons with the time constant $\kappa_A$=$\kappa_B$=1.5.
Details of the simulation are given in Sec.II.E and Table \ref{table2} in Appendix.
}
\label{Fig6}
\end{figure}

For the sumulation of the canonical model of a class 1 neurons(Fig. \ref{Fig6}), 
All the parameters are tabulated in Table \ref{table2}.

\begin{table*}[htbp]
\caption{Parameters for the numerical simulation of simple threshold units shown in Figs. \ref{Fig2}d-f and \ref{Fig4}a-c.}
\begin{tabular}{|c||r|r|r|r|r||r|}
\hline
	& \multicolumn{5}{c||}{numerical simulations} & elec.circuit \\
\hline
figure number	& 2d & 2e and f & 4a & 4b & 4c & 3b\\
\hline
\hline
$M$	& \multicolumn{5}{c||}{4} & 4 \\
\hline
$N$	& 1 -- 1000 & 100 & 1 & 1 -- 100 & 100 & 1\\
\hline
$V^{\rm Th}_m$ &	0.10	& 0.20 & - & -	& - & - \\
\hline
$\epsilon V_m$ &  0.16	& 0 -- 0.50 & - & - & - & - \\
\hline
$t_{\rm pw}$ & \multicolumn{2}{c|}{128} & - & - & - & - \\
\hline
$V^{\rm Th}_A$ & - & - & \multicolumn{3}{c||}{0.05} & 0.03V \\
\hline
$V^{\rm Th}_B$ & - & - & \multicolumn{3}{c||}{0.20} & 0.06V \\
\hline
$\epsilon V^A_m$ & - & - & \multicolumn{3}{c||}{0.30} & 0.16V \\
\hline
$\epsilon V^B_m$ & - & - & \multicolumn{3}{c||}{0.30} & 0.25V \\
\hline
$\omega/2\pi$ & \multicolumn{4}{c|}{6.25$\times 10^{-2}$(=1/16)} & 10$^{-5}$ -- $10^0$ & 0.3Hz\\
\hline
$D$ & 0.10 & 0 -- 0.50 & 0.05 -- 0.13 & \multicolumn{2}{c||}{0.11} & 1.6V -- 2.8V\\
\hline
$\tau$ & \multicolumn{5}{c||}{16} & - \\
\hline
time-step & \multicolumn{5}{c||}{1} & 1$\times 10^{-3}$s \\
\hline
averaging time & \multicolumn{5}{c||}{64} & 0.1s \\
\hline
\end{tabular}
\label{table1}
\end{table*}

\begin{table*}[htbp]
\caption{Parameters for the numerical simulation shown in Fig. \ref{Fig6}.}
\begin{tabular}{|c||r|r|r|}
\hline
figure number & \hfill{} 6a & \hfill 6b & \hfill 6c \\
\hline
\hline
$M$ & \multicolumn{3}{c|}{4}\\
\hline
$N$ & \hfill 50 & \hfill 1,4,10,25,50 & \hfill 50 \\
\hline
$r_A$ & \multicolumn{3}{c|}{-0.008}\\
\hline
$r_B$ & \multicolumn{3}{c|}{-0.020}\\
\hline
$\epsilon_{ex}^A$ & \multicolumn{3}{c|}{0.5}\\
\hline
$\epsilon_{ex}^B$ & \multicolumn{3}{c|}{0.5} \\
\hline
$\epsilon_{inh}$ & \multicolumn{3}{c|}{3.0}\\
\hline
$A_{\rm sig}$ & \multicolumn{3}{c|}{0.019}\\
\hline
$\omega / 2\pi$ & \multicolumn{2}{c|}{1.0$\times 10^{-3}$} & $2.0\times 10^{-4}$ -- $5.0\times 10^{-3}$ \\
\hline
$D$ & \hfill 6.0$\times$10$^{-4}$ -- 1.4$\times$10$^{-3}$ & \hfill 1.0$\times$10$^{-3}$ & \hfill 1.0$\times$$10^{-3}$  \\
\hline
$\kappa_A$ & \multicolumn{3}{c|}{1.5}\\
\hline
$\kappa_B$ & \multicolumn{3}{c|}{1.5}\\
\hline
time-step & \multicolumn{3}{c|}{0.05}\\
\hline
averaging time & \multicolumn{3}{c|}{50}\\
\hline
\end{tabular}
\label{table2}
\end{table*}

%\newpage
\clearpage

\clearpage

\end{document}